# Refined HLA Linkage Disequilibrium Architectures of World Populations by a Novel Allelic Correlation Measure

*Fei Zhang*[1,2] and *Weixiong Zhang*[1,2,3,4*]

1: Hong Kong Jockey Club STEM Laboratory of Genomics and AI in Healthcare,
2: Department of Health Technology & Informatics,
3: Department of Data Science & Artificial Intelligence,
4: Department of Computing,
   The Hong Kong Polytechnic University, Hong Kong SAR, China

*. Correspondence: WZ: weixiong.zhang@polyu.edu.hk

**Abstract**

Numerous diseases, particularly autoimmune disorders, are associated with the human leukocyte antigen (HLA), a small genomic region located on human chromosome 6. Adequate characterization of linkage disequilibrium (LD) in the HLA across populations is crucial for identifying genetic markers associated with specific traits and phenotypes. However, current LD measures often fail to capture HLA's structural complexity due to methodological limitations and sensitivity to low-frequency variants, marginal allele frequencies, and haplotype composition. To address these challenges, we introduced the Conditional Informatics Correlation Coefficient (CICC), which integrates conditional probability, information content, and haplotype-aware XOR logic to quantify LD robustly. When applied to high-resolution haploid genomes from the Human Pangenome Reference Consortium (HPRC), CICC revealed 10 novel high-LD regions in HLA. Further analyses using the 1000 Genomes Project and Genome Asia datasets identified nine strongly linked regions shared across five global populations—five in Class I and four in Class II. These results demonstrate CICC's ability to capture complex HLA LD structures across populations, highlighting its broad potential for disease gene mapping, population genomics, and guiding precision medicine.

*Keywords: linkage disequilibrium, human leukocyte antigen, CICC, haplotype genome*

## 1. Introduction

The human leukocyte antigen (HLA) on human chromosome 6, also known as the major histocompatibility complex (MHC) for animals, is one of the most genetically heterogeneous and gene-rich regions in the human genome[1]. Numerous studies have drawn attention to HLA genetic polymorphisms, many of which are linked to various diseases[2], such as autoimmune diseases like psoriasis[3,4] and major psychiatric disorders (MPDs), including depression, schizophrenia and bipolar disorder[5–7]. These medical complications are polygenic, influenced by multiple genotypic variations of single-nucleotide polymorphisms (SNPs)[8,9]. The genetic interactions or correlations among multiple variation loci, some of which appear as epistasis and pleiotropy[10], are often the main contributors to disease etiologies[11]. Understanding the correlations among the variants within the HLA can help reveal pathogenic mechanisms crucial for treating HLA-associated diseases.

It is critically important to quantify the SNP correlations in relation to the underlying linkage disequilibrium (LD) architecture because LD captures non-random associations among loci[12]. While efforts have been made to characterize the human genome's LD architecture[13–15], a definitive LD map remains elusive. Detailed LD maps exist for chromosomes 3 and 19[16], but not for chromosome 6—particularly for the intricate HLA—despite its importance to human genetics and medical implications. Existing LD structures for the HLA are inconsistent: HLA-Cw/B blocks in HLA class I among British individuals can extend to HLA-A[17]. Similarly, HLA-DR/DQ blocks in HLA class II for both British individuals and Ashkenazi Jews can extend to HLA-DPB1[18,19]. Additionally, a tumor necrosis factor (TNF) block containing TNF-α and TNF-β has been reported to be present within HLA class III[17]. These incomplete and inconsistent findings underscore the need for a unified and accurate delineation of the LD structure across the HLA region and the whole genome. These findings also call for novel approaches to accurately identifying LD architecture.

In the absence of accurate fine-scale LD maps, most existing studies approximate LD architecture by partitioning population-level genotype data [13,16,20] into haplotype blocks (HBs)[21–25]—regions of highly correlated variants that tend to be inherited together. Popular methods for identifying HB boundaries include the confidence interval (CI)[26], solid spine (SS)[27], and four-gamete test (FGT)[28], which are implemented in tools such as Haploview[27] and PLINK[29]. While widely adopted, these block-based methods often lack the resolution needed to capture the true underlying allelic correlations. Their inferred boundaries are sensitive to population structure, allele frequencies, and algorithmic assumptions, resulting in considerable variability across studies.

Accurate LD structures identification requires precise haplotype phasing information and appropriate LD measures. Advances in the T2T-CHM13 reference genome[30,31] and the widespread adoption of long-read sequencing technologies[31] have enabled the direct acquisition of high-precision haplotypes, avoiding errors inherent in traditional phasing. The widely adopted frequency-based LD measures such as $r^2$[32], $D'$[33] and allele-based Custom Correlation Coefficient (CCC)[34] are common, but $r^2$ and $D'$ treat SNPs as atomic units, unlike CCC. These measures are perplexed by three serious issues. (1) $r^2$ and $D'$ heavily rely on allele marginal probabilities, yielding identical values for SNP pairs with the same marginal probabilities but different joint distributions. (2) Low-frequency variations are typically excluded that may be disease-associated[35–39]. (3) These measures ignore the haplotype relationship among SNPs by treating maternal and paternal chromosomes as a single, hybrid entity, disregarding phasing information crucial for genetically heterogeneous populations[40,41]. Indeed, allele-specific expression (ASE) highlights the importance of phasing among multiple loci in pinpointing the causal genetic variations in disease diagnosis[10].

To address these limitations, we developed a novel haplotype-aware LD measure named Conditional Informatics Correlation Coefficient (CICC). CICC effectively mitigates the influence of instability factors on SNP LD strength and incorporates haplotype information to improve accuracy and robustness. We validated CICC's performance using simulated and real population data. CICC detected 10 novel strongly correlated regions based on haplotype genomes[31]. Applying CICC to the 1000 Genomes Project (1000G)[42] and Genome Asia Project (GAsP)[43] datasets, we identify nine strong haplotype blocks in the HLA that are shared among European, African, American, and Asian populations. These refined LD maps provide a detailed view of population-specific architecture and migration histories, offering a valuable resource for downstream genetic analyses and clinical applications.

## 2. Result

We developed a novel LD measure, CICC, to construct LD structures (Methods; Appendices I - III). By utilizing CICC in conjunction with genomic sequences from HPRC[31], 1000G[42], and GAsP[43], we constructed population-specific LD structures. Our analysis identified ten novel regions with high LD, as well as shared LD structural patterns across diverse populations (Fig. S1).

### 2.1. Conditional Informatics Correlation Coefficient (CICC)

The popular existing LD or correlation measures, including $r^2$, $D'$ and CCC (Appendix IV), share three limitations when applied to LD detection. Firstly, they fail to incorporate the phasing information of SNPs, i.e., the pairing pattern of adjacent SNPs, which is essential for capturing inheritance structure. Secondly, they are overly dependent on marginal allele frequencies, treating them in isolation and often ignoring joint base distributions. This simplification obscures intricate combinatorial patterns, particularly in the presence of skewed allele frequencies. Thirdly, rare variants are often excluded from the analysis, while in reality, although rare, they can play an important role in interpreting the causal relationships between SNPs[36,38]. As a result, LD results are obscured.

Therefore, we introduced a novel LD measurement called CICC. We summarize the core formulation of CICC below and leave its details to the Methods section and Appendix I.

$$CICC = max(CICC_{ij}) = \max\left(\Phi(X,Y) * IF_{(i|j)} * IF_{(j|i)}\right), \; i \in \{A, a\}, j \in \{B, b\}$$

where

$$\begin{cases} \Phi(X,Y) = \frac{1}{2K} * \sum_{\substack{n=1 \\ g(x_n) \oplus g(y_n)=0}}^{2K} 1, g(t) = \begin{cases} 0, & t = REF \\ 1, & t = ALT \end{cases} \\ IF_{(i|j)} = \frac{2}{\pi} * \arctan\left(\frac{1}{-\log_2(P_{(i|j)})}\right) \end{cases}$$

In the formula, $X$ and $Y$ denote two loci or SNPs over a population of K individuals, with possible alleles "A"/"a" and "B"/"b", respectively, so there are $2K$ total alleles per SNP across the K individuals. Let $x_n$ and $y_n$ be the $n$th allele for X and Y ($n = 1, 2, \ldots, 2K$). $\Phi(X,Y)$ quantifies allelic concordance between $X$ and $Y$ by evaluating whether both alleles conform to the same state (REF or alternate allele (ALT)) using an XOR function. The Informativeness Factor (IF), based on conditional probabilities, accounts for skewed allele frequencies and preserves information from rare variants.

As detailed in Appendices I-III (Fig. S2-S5 and Table 1A-C; 2A; S1-S3), CICC addresses these challenges through three key features. The first and most imminent is its capability of integrating haplotype information in quantifying pairwise correlations, which is critical in capturing accurate LD structures and important for disease studies[41,44,45]. This feature becomes increasingly important as more haplotype genomes have been sequenced recently[30,31], and this trend is expected to continue, making genetic studies robust. CICC incorporates the allelic concordance between two genomic loci or SNPs in the $\Phi(X,Y)$ function (Appendix I). The second feature is that it can retain low-frequency variants and capture their contributions by using conditional information content with an arctan-based nonlinear transformation in the IF function. The third feature is its capability of subduing the effect of marginal probabilities by using $\Phi(X,Y)$ to ensure that distinct allelic arrangements with the same marginal probabilities are not erroneously treated as equally correlated.

The following two applications, with some of the ground truth available, illustrate the rigor and utility of CICC in capturing LD structures and complex disease studies.

### 2.2. CICC accurately recovers the validated recombination hotspots using the 1000 Genomes Project.

A recombination hotspot is a region where recombination occurs frequently. As a result, the LD between SNPs within this region is typically very weak[35]. A reliable LD metric should be able to accurately identify hotspot regions by delineating highly correlated regions into blocks. Thus, we compared and analysed eight combinations of four LD measurement methods (i.e., $r^2$, $D'$, CCC and CICC) and three methods (including CI, SS and FGT) for identifying haplotype blocks (Fig. 1A). The HLA contains six distinct crossover hotspots – DNA1, DNA2, DNA3, DDM1, DDM2, and TAP2 (Fig. 1A) – that are validated by sperm cross-experiments[35].

An effective LD measure combined with an appropriate block partitioning strategy can identify recombination hotspots. In our study, we applied this integrated approach (Section 4.2) to the HLA, and the resulting LD blocks showed strong concordance with known recombination hotspots based on direct overlap analysis (Appendix III).

An effective LD metric should also be able to distinguish recombination hotspots from regions of high linkage, as the correlation distributions in these regions are expected to differ. To validate this, we applied the Kolmogorov-Smirnov (KS) test[46] (Fig. 1B) and Kernel Density Estimation (KDE)[47] (Fig. 1C-E; S6). These analyses demonstrated that CICC more effectively captures the differences in distribution patterns within recombination hotspot regions compared to existing metrics (Appendix III).

### 2.3. CICC reveals greater GWAS enrichment than $r^2$ across multiple diseases.

LD measurement is central to population history research and genome-wide association studies (GWAS). Typically, analysis begins by identifying core disease-associated loci via statistical models[38,48], followed by applying LD metrics to detect neighboring loci in strong correlation. This expands the candidate set for downstream functional and mechanistic studies

We postulated that disease-associated loci exhibit elevated LD in affected individuals but reduced LD in unaffected ones, likely due to allelic variation disrupting cooperative interactions—a hypothesis supported by multiple studies[49–51].

Guided by this, we analyzed genotype data from psoriasis, Alzheimer's disease (AD), and psychiatric disorders (PD), focusing on chromosome 6, a region implicated in immune and neuropsychiatric traits. We compared CICC and $r^2$ in recovering GWAS Catalog–validated loci (Appendix III). CICC consistently identified a higher proportion of validated loci (Fig. 1F), indicating increased sensitivity and improved interpretability. In psoriasis, CICC outperformed $r^2$, capturing 17.1% and 8.2% of Found variants vs 16.9% and 0 % by $r^2$, in the two psoriasis datasets, respectively. For AD, CICC captured 1% of Found variants, while $r^2$ captured none (0%) (Fig. 1F), thus identifying more relevant variants and achieving higher validation rates. For PD, both performed similarly, reflecting greater genetic complexity and the need for a larger cohort or more refined analyses (Table 2B).

### 2.4. CICC identifies ten novel LD-linked regions in the genome

CICC excelled in haplotype-based genome analysis by integrating phasing information and conditional probability. It precisely delineated regions of strong linkage both within and between genes, offering deeper insights into genomic structure and haplotype organization. Within the highly complex HLA region, CICC showed notable accuracy in detecting SNP-level LD patterns.

Beyond recovering most known gene boundaries (Appendix III, Fig. 2), CICC discovered ten previously unrecognized regions of strong LD within the HLA—three in Class I and seven in Class II—reflecting its heightened sensitivity to intricate LD architectures. Several of these novel regions exhibited strong correlations with established loci, revealing previously undetected, latent layers of organization in the HLA and offering new avenues for functional and evolutionary investigation (Table 2C).

### 2.5. LD structures of the HLA across five world populations

Gene flow and environmental factors influence the prevalence of certain gene variants, thereby affecting the diversity of HLA and enhancing population adaptability and disease resistance[52,53]. The HLA region is subjected to more profound environmental selection pressures. What, then, are the true LD structures of the HLA across populations? To answer this challenging question, we conducted a population-level analysis at the superpopulation scale to investigate the interplay between genetic architectures and population structure, and systematically established region-specific LD patterns tailored to each population.

### 2.5.1. HLA Class I

We identified five novel high-LD regions in HLA Class I across different populations (EAS, SAS, AMR, AFR, EUR) (Fig. S7), each exhibiting distinct population-specific LD profiles.

Region 1 (HLA-F to HLA-V): AFR exhibited the strongest LD, with EUR and Asian populations also showing extended correlation into ZFP57. EAS showed significantly stronger LD than SAS; SAS and AMR displayed comparable patterns. Region 2 (HLA-W to RPP21): This region displayed strong LD and could be partitioned into three subregions: (i) HLA-W to TRIM31 (correlated to Region 1), (ii) TRIM40 to TRIM26, and (iii) TRIM26BP to RPP21.

Regions 3 & 4 (TMPOP1 to IER3; HCG20 to HCG21): AFR showed the highest LD, followed by EAS and AMR; EUR showed markedly weaker LD, reflecting regional divergence. Region 5 (DHFRP2–HCP5): AFR showed strong LD, while AMR was intermediate between EAS and SAS. EAS appeared more conserved than SAS. LD dropped notably at HCP5, especially in EUR.

While population-specific differences were evident, shared trends emerged. AFR consistently exhibited larger and more contiguous LD blocks, followed by Asia. EAS showed higher block correlations than SAS, resulting in larger blocks. AMR showed increased fragmentation, and EUR displayed the most disrupted structure, suggesting higher haplotype diversity and recombination. Notably, Regions 3 and 4 exhibited strong inter-regional correlation, as did the remaining three regions.

### 2.5.2. HLA Class II LD Patterns

Several key features were identified in the HLA Class II region (Fig. S8). A clear boundary between HLA-DRA and HLA-DRB9 was observed in all populations except AFR. In Asians, lower LD in the HLA-DRB9 region was partly due to limited SNP coverage in GAsP samples. Within Asia, SAS showed stronger internal LD than EAS.

Region 1 (TSBP1 to HLA-DRA): This region exhibited high LD across all populations and was strongly linked to regions 2 (HLA-DQA2 to HLA-DQB2), 3 (HLA-DOB to HLA-DOA) and 4 (HLA-DPA1 to HC24). The highest correlation between TSBP1 and HRNPA1P2 within Region 1 was observed in AFR, followed by Asian and AMR. Within Asian, EAS and SAS showed similar levels, slightly below AMR, with EUR the weakest.

Region 3 (gene HLA-DOB to HLA-DOA): The overall structure of LD in Asians was similar to that of EAS. SNP correlations in Asians and EAS were both greater than in SAS. The correlation level in SAS was approximately equal to that in AMR, and both were higher than in EUR.

Region 4 (gene HLA-DPA1 to HCG24): The correlation between SAS and HLA-DPA1 was higher than that between AMR. In SAS, the differentiation between HLA-DPA2 and COL11A2P1 was more pronounced. For the EUR, this entire region exhibited the weakest correlation of all regions. Asian was more similar to SAS, while EAS was more similar to the correlation structure of EUR. The overall order was: Asian≈SAS>EAS>AMR>EUR.

### 2.5.3. HLA Class III

Our genetic structure analysis of the HLA class III region revealed a significant presence of high LD, indicating less distinct block partitioning compared to classes I and II (Fig. S9A-F). This phenomenon is primarily influenced by two factors. Firstly, despite being in a high overall recombination rate zone within the HLA, class III exhibits local recombination cold spots in gene-dense areas. These cold spots enable the stable inheritance of SNPs and alleles across generations.

Secondly, the critical role of the HLA region in immune response has subjected it to strong selective pressure. Positive or stabilizing selection mechanisms tended to preserve advantageous gene combinations, establishing a robust LD. Notably, the HLA class III exhibits the highest gene density among the three classes (gene density: 8.5E-05; 60 genes / ~0.7Mbp), potentially amplifying the impact of selection pressure compared to class I (4.69E-05; 163 genes / ~3.5Mbp) and class II (3.72E-05; 32 genes / ~0.86Mbp). Simultaneously, genetic drift played a pivotal role in small or geographically isolated populations by randomly altering allele frequencies, further intensifying LD.

Consequently, these factors contributed to the distinct, yet globally consistent LD architecture observed in HLA Class III regions.

*2.5.4. Summary*

The nine highly correlated regions spanning HLA class I and II were categorized by LD strength. Linkage gradually decreased from the AFR to the EUR, resulting in increased fragmentation within the highly related haplotype blocks. In contrast, HLA class III showed strong and consistent LD patterns across populations. Overall, LD was strongest in AFR and weakest in EUR, with notable differences between EAS and SAS (Fig. S9G).

African populations showed significantly longer and more stable LD blocks than others, followed by Asians and Americans, while Europeans had the most fragmented structure. This pattern contrasts with the general genome-wide trend of LD decay in Africans driven by higher recombination rates, indicating that selective pressures on functionally important regions strongly shape LD architecture[54].

The refined HLA genetic structure clearly demonstrated a gradual differentiation in regions with high HLA correlation—from Africa to Asia, from the Americas to Europe—accompanied by a reduction in the number of high-correlation blocks at the individual level. Shared patterns across populations likely reflect historical migrations and gene flow, preserving common genetic features, whereas distinct regional differences appear to be driven by local environmental pressures shaping adaptive variation in HLA genes. Together, these findings highlight how natural selection and environment shape population-specific LD in immune-related regions.

**2.6. Comparison with current LD structures**

Previous studies have identified three major LD blocks: the HLA-C/B block (class I), the HLA-DR/DQ block (class II), and the TNF block (class III)[17–19,55]. In the HLA-C/B region, we observed strong LD signals within the HLA-C and HLA-B genes themselves, while the correlation between these two genes was relatively weak (Fig. 2A).

The conserved HLA-DR/DQ block, confirmed in multiple populations[18,19,55], extends to include HLA-DPB1 in both the British population and Ashkenazi Jews [18]. However, in the LD analysis of German population data[56], strong correlations were observed only within HLA-DRA and HLA-DQA1, while no significant LD was detected between these two loci[56].

In this study, leveraging high-quality haplotype data from HPRC, CICC resolved this block (Fig. 2E), whereas unphased datasets (1000G, GAsP) failed to capture it reliably. These findings demonstrate that CICC is effective in delineating fine-scale and reliable LD structures by incorporating haplotype information, and it detects both local and long-range LD patterns that are overlooked by conventional metrics (Fig. S10). Additionally, CICC successfully identified the TNF block in the HLA class III region, encompassing the TNF-α and TNF-β genes (Fig. S9) [17,56].

It is worth noting that the existing whole-genome LD maps have lacked resolution in the HLA, producing coarse partitions inconsistent with classical HLA classes[13,21,57]. Through the accurate reconstruction of detailed LD structure maps, we can identify more precise and biologically meaningful haplotype architectures. The CICC method, especially when applied to high-quality haplotype data, excels in this task, providing a robust framework for fine-scale genetic analyses and downstream applications.

**3. Discussion**

In this study, we address a challenge in human genetics: the lack of a high-resolution, population-wide map of LD within the HLA region. To fill this gap, we introduced CICC, a novel LD metric that fully leverages haplotype-resolved genomes while addressing key limitations of existing methods. Using CICC, we constructed the first detailed LD atlas of the HLA across global populations, revealing both previously uncharacterized and evolutionarily conserved LD structures. Beyond providing a methodological advance, the resulting maps reveal the true organizational principles of this highly polymorphic locus, resolving both conserved and population-specific haplotype structures. These insights redefine our understanding of how genetic variation in the HLA shapes immunity, disease susceptibility, and human diversity, and establish a foundation for future work in disease mapping, evolutionary genetics, and translational research.

CICC identified more functionally validated variants from the GWAS catalog than $r^2$, suggesting that conventional LD measures may underestimate pathogenic regions and incompletely resolve causal haplotype blocks. This limitation may partly explain the modest success of GWAS-driven therapies, as incomplete mapping overlooks critical disease loci. By incorporating haplotype structure and rare variants, CICC enables the identification of functional variants that are overlooked by conventional LD measures, as evidenced by the greater enrichment of GWAS-catalog–validated loci. It offers a more direct path to disease mechanisms and therapeutic targets. At the same time, the refined LD structures revealed by CICC—including strong associations in regions such as HLA-DR/DQ—align closely with functional units and uncover population-specific patterns shaped by migration and selection. Together, these findings establish CICC as a reliable framework for advancing both disease genetics and evolutionary inference.

At the population level, the high-resolution LD atlas generated by CICC recovered both conserved structures and novel high-LD segments. African populations exhibited the most extensive and stable haplotype structures, East Asian populations showed more conserved LD consistent with gradual northward migration and geographic isolation[58,59], while South Asian populations displayed weaker LD[60], reflecting complex waves of migration. These findings illustrate how haplotype-resolved LD maps can provide a more accurate basis for inferring human origins, migration routes, and population-specific disease risks.

While this study provides valuable insights, there are several aspects that remain to be addressed. CICC relies on phased haplotype data, which remain challenging to obtain at a population scale despite advances in long-read sequencing, and current algorithms cannot yet guarantee fully accurate reconstructions. Our analyses also focused on correlations without explicitly modeling epistasis or causal mechanisms, limiting mechanistic interpretation. Future work will integrate CICC with causal inference frameworks, extend analyses to the human pangenome, and combine with multi-omics data to provide deeper insights into disease mechanisms and human evolution.

## 4. Methods and Materials

### 4.1. The new measurement - conditional informatics correlation coefficient (CICC)

CICC comprises two components: a genetic compatibility factor and an information factor (Appendix I and Fig. S11). The compatibility factor quantifies how the alleles of two genomic loci are compatible over the individuals examined. Note that it utilizes alleles, thereby incorporating phasing information and accounting for the mutation status of each allele. The compatibility factor is implemented in an XOR function using the REF in the reference genome. The alternate allele is the mutated allele (ALT).

The provided information indicates the presence of mutations in each base, as well as whether they originate from the paternal or maternal allele. Next, it assesses whether each SNP pair displays a comparable distribution across haplotype genomes: REF homozygous or ALT homozygous pairs are marked as 1, while REF and ALT heterozygotes are marked as 0. The relevant formula is presented below:

$$\Phi(X,Y) = \frac{1}{2K} * \sum_{\substack{n=1 \\ g(x_n) \oplus g(y_n)=0}}^{2K} 1, \quad g(t) = \begin{cases} 0, & t = REF \\ 1, & t = ALT \end{cases} \quad (1)$$

$X$ and $Y$ denote two SNPs of each pair of SNPs. $x$ represents the bases of SNP $X$, and $y$ represents the bases of SNP $Y$, and $g(x)$ represents the conformity of each base with the reference allele - REF (0 for consistent, 1 for inconsistent). $n$ represents the base number in a bi-allelic SNP. For example, for the SNP C|G, $n = 1$ represents C and $n = 2$ represents G. $x_k$ denotes the consistency of base $x$ in the $k$th individual's $X$ SNP allele. $y_k$ denotes the consistency of base $y$ in the $k$th individual's $Y$ SNP allele. There are $K$ individuals, and each person has two bases, so there are $2K$ bases. The value for each individual SNP pair is 1 when both SNPs conform to either the reference or alternate allele - ALT (Table 1D). When two SNPs are completely in frequency synchrony (i.e., both SNPs are conformed with REF alleles), $\Phi(X,Y)$ will equal 1. When the variations of the two SNPs are completely asynchronous, $\Phi(X,Y)$ will equal 0.

Secondly, the Information Factor (IF) integrates conditional information to regulate the impact of low-frequency and rare variants while quantifying causal dependence between alleles in SNP pairs. To enhance the robustness of LD structure analysis, researchers often exclude these variants. However, low-frequency and rare variants have been shown to significantly influence LD, especially in complex diseases[36–38,61]. Due to their higher stochastic bias compared to common alleles, IF uses information content to capture the effects of all variations, including rare ones. In information theory, higher uncertainty (lower probability) corresponds to greater information content, while lower uncertainty (higher probability) yields less information[62].

Moreover, the conditional amount of information is utilized to quantify the portion in one base that cannot be explained or predicted by another base, thereby estimating the degree of causal dependency between the two bases. The core part is $P_{(i|j)}$, which is employed to characterize an association among SNPs, and such a relationship can be leveraged to estimate and predict the behavior of another gamete or allele. Assuming there is an allele $j$ in another SNP, the probability of observing an allele $i$ in a SNP under these circumstances is denoted as $P_{(i|j)}$. Estimation of allele probabilities is accomplished by treating the two haplotypes within an individual as independent samples. This approach enables the assessment of allele frequencies derived from haplotype information. By integrating data from different haplotypes, it yields a more precise depiction of the population's nucleotide distribution.

The sum of joint probabilities ($P_{ij}$) for allele pairs at each SNP equals 1, where $i$ and $j$ represent the alleles at two SNP loci. For example, $A$ and $a$ are alleles at SNP1, while $B$ and $b$ are alleles at SNP2 (Table 1B). Thus, we have $P_{AB} + P_{Ab} + P_{aB} + P_{ab} = 1$. Apart from the enhancements in rare and low-frequency variations, IF cannot only quantify causal dependencies by employing conditional probabilities rather than direct probabilities of alleles, but also diminish the reliance on marginal probabilities.

Owing to the influence of the value range and the log function, an infinite value emerges when the conditional probability approaches 1. To render the correlation comparable and preserve the nonlinear variations, the inverse arctangent ($arctan$) function is selected for the nonlinear transformation of the function. To intuitively represent the degree of linkage within the range of 0 to 1, we incorporate the factor $\frac{2}{\pi}$. Given that x belongs to the real number domain, the value range of the inverse trigonometric function $arctan(x)$ is $(-\frac{\pi}{2}, \frac{\pi}{2})$. Consequently, the feasible values for IF fall within $(0,1)$. For values approaching 0 and 1 infinitely, take 0 or 1 in the calculation.

$$IF_{(i|j)} = \frac{2}{\pi} * \arctan\left(\frac{1}{log_2\left(\frac{1}{P_{(i|j)}}\right)}\right), \ i \in \{A, a\}, j \in \{B, b\} \tag{2}$$

The final CICC formula is shown as follows:

$$CICC = max(CICC_{ij}) = \max\left(\Phi(X,Y) * IF_{(i|j)} * IF_{(j|i)}\right), \ i \in \{A, a\}, j \in \{B, b\} \tag{3}$$

Two SNPs yield four possible allele combinations, each pair inherited maternally or paternally. Chromosomal base correlations remain strong despite gene mutations or regions with low recombination rates; when two sites form a haplotype block due to high correlation within one chromosome, their counterparts in another chromosome also constitute a haplotype block by nature. Henceforth, we adopt the maximum value among these combinations as an aggregate measure for correlation assessment.

CICC serves as a novel LD measurement method with several distinctive features: 1) It liberates itself from assumptions regarding data distribution, enabling applicability across linear and nonlinear datasets. 2) The integration of conditional probability facilitates exploration of potential causal relationships between SNPs while augmenting inter-SNP correlation identification through combined prior probabilities. 3) The incorporation of information content principles effectively regulates impact evaluation for low-frequency variants (MAF<5%). 4) The computation process integrates considerations for allele consistency and heterozygosity within populations, thereby fortifying resilience against rare variants and heterozygotes. 5) Diverse statistical functions minimize dependence on marginal probabilities.

**4.2. Haplotype partitioning methods**

Haplotype blocks contain highly correlated genetic variants that tend to be inherited together. Several factors, including the choice of LD measures, haplotype frequencies within sampled populations, and the extent of chromosomal mutations and recombination events influence the determination of HB boundaries. Popular methods for identifying HB boundaries include the confidence interval (CI)[26], solid spine (SS)[27] and four-gamete test (FGT)[28]. The CI method identifies strongly correlated SNP pairs by establishing a 95% confidence interval on the values of $D'$, a prevalent LD measure[26]. The SS method determines block boundaries based on the log odds of base ratios for $D'$ and progressively refines block boundaries[27]. The SNP loci within these boundaries—referred to as spine ranges—are considered to belong to the same haplotype block. Assuming no recombination within haplotype blocks, FGT estimates haplotype blocks based on recombination rates based on the probabilities of haplotypes from four gametes of each pair of bi-allelic SNPs[28]. These HB boundary detection methods have been incorporated into broadly adopted LD analysis tools such as Haploview[27] and GWAS pipelines like PLINK[29]. Haploview can adopt the CI, FGT, or SS independently, and PLINK employs the CI in conjunction with the popular LD measure $r^2$.

### 4.3. Subjects and data collection

This study examined a total of 11,179 samples, comprising 2,984 samples derived from simulated genetic data and 3,202 samples obtained from the third phase of the 1000 Genomes Project Consortium (1000G)[42], in addition to 1,163 samples sourced from the Asian Genome 100K project (GAsP)[43] and 47 individuals from the Human Pan-Genome Reference Consortium (HPRC)[31]. And 3,783 individuals from disease analysis. Furthermore, it identified a total of 94 haplotypes[31] through whole-genome sequencing. The primary objective of the comprehensive analysis undertaken by the 1000G is to catalog prevalent genetic variants within human populations with a minimum frequency of at least 1%, encompassing SNPs and structural variants.

To accurately assign genotypes to all identified variation sites, data from diverse populations were integrated into this project's dataset, consisting of over three thousand samples across 26 populations categorized into five superpopulations: East Asian (EAS), South Asian (SAS), American (AMR), African (AFR), and European (EUR). Additionally, the pilot phase of GAsP includes reference datasets for whole-genome sequencing, representing individuals from various populations in 64 countries across Asia. Following filtration processes, a subset comprising 1163 representative samples was selected.

For these samples, our analysis focuses on the HLA region (GRCh38 coordinates chr6:28,510,120-33,480,577)[42]. The HLA region, essential for immune response, exhibits significant genetic variation[63–65]. During data processing, we utilized positional information on chromosome 6 to isolate the HLA region from all provided samples. Exclusion criteria encompass: 1) reference or variant alleles with a non-unit length; 2) duplicate records; 3) SNP IDs and positions with missing values on the chromosome; and 4) incomplete SNP coverage across all samples in the region. 4) Due to the lack of consistency between CCC values and other methods, self-correlation calculation results were excluded from gene and haplotype analysis to ensure methodological coherence.

For the disease enrichment analysis, we utilized five distinct disease datasets, with a specific focus on chromosome 6. We analyzed two psoriasis datasets obtained from dbGaP (accession number: phs000019.v1.p1.c1), accessed under General Research Use (GRU) and Autoimmune Disease Only (ADO) permissions. The GRU dataset originally comprised 451,724 SNPs across 1,683 individuals. Following the exclusion of children and participants who failed the original study's quality control criteria[66–68], we retained only variants located on chromosome 6. After applying our predefined filtering criteria, a total of 30,875 high-quality, non-missing SNPs remained. Among the filtered cohort, 929 individuals were diagnosed with psoriasis, and 681 served as unaffected controls.

The ADO dataset included genotypes at 451,724 loci for 1,214 individuals. After removing individuals who did not pass quality control, 1,167 samples remained. Restricting the analysis to chromosome 6 and applying the same filtering criteria yielded 30,670 SNPs with complete data. This final dataset consisted of 439 psoriatic cases and 728 controls.

Genotype data for Alzheimer's disease (AD) were acquired from the Alzheimer's Disease Neuroimaging Initiative (ADNI) (https://adni.loni.usc.edu/), involving 785 individuals. We restricted the study to chromosome 6 and employed the identical quality control and filtering criteria previously outlined. The total dataset had 926,956 SNPs, all of which were complete with no missing values. Among the retained persons, 527 were clinically diagnosed with Alzheimer's disease, whereas 258 were cognitively normal controls.

Genomic data pertaining to psychiatric diseases were obtained from the NCBI Sequence Read Archive, namely under accession numbers PRJNA551447. The PRJNA551447 sample comprised 101 unrelated Han Chinese people, consisting of 49 schizophrenia (SCZ) patients and 52 unaffected controls.

**4.4. Population genetic simulation algorithm**

The population genetic simulation algorithm, denoted as Algorithm 1 (Appendix V), is capable of simulating gene sequence dynamics during population iteration and inheritance. Following the provision of the initial set of gene sequences, a restricted series of genetic iterations is executed, which includes gene recombination, genetic mutation, and various other genetic phenomena. A "gene screening" methodology is utilized during this procedure to select individuals for population updates. The primary goal of simulating genetic regions that exhibit linkage disequilibrium is to validate the correlation between SNPs. The simulation algorithm's output includes data about SNPs and the detection of regions distinguished by LD. The data provided serves as a verification tool for the computation of LD. Therefore, the initial gene recombination process is impacted by the simulation program through the intentional establishment of regions with high correlation and low correlation. Prophase I of meiosis is the site of gene recombination, which occurs when non-sister chromatids on homologous chromosomes cross over during synapsis. This crossing constitutes the principal determinant of linkage disequilibrium. $N$ has been set as 200, $G$ as 50,000, and $M$ as 360.

Gene mutation is a process in which every gene position in a population is stochastically modified in accordance with a pre-established probability. Region generation: To begin, generate a collection of linkage disequilibrium regions, with each region comprising three position segments. These regions must adhere to the following criteria: recombination must involve the synchronous exchange of all positions within the three segments, and none of the subregions may overlap or cross the boundaries. Then, establish two recombination hotspot zones, wherein positions with various lengths and unlimited initial positions are permitted to exchange bases. Strict adherence to the legal length restrictions imposed on the gene sequence is of utmost importance for all region settings.

As described in Algorithm 2 (Appendix V), gene recombination begins with the initial random selection of individuals from a population's memory pool. The following requirements must be met by these individuals: they cannot share any common ancestors within the past three generations, and their lifespans must span at least one to three generations. One to three children will be generated at random for each selected pair of parents. Determining gene recombination throughout the genetic process is crucial in numerous locations. When recombination does occur, it is crucial that recombination operations on gene positions are carried out in accordance with region-specific laws. With the incorporation of offspring genotypes, the population's memory pool list is ultimately revised.

Individual members of the population should be updated. Individuals who meet the following criteria should be excluded from both the roster of existing population IDs and the genotyping array of the population: When the lifetime reaches five and the population size exceeds 20 times its initial quantity, 20%–50% of the individuals exhibiting the greatest deviation from the "optimal genes of the population" should be removed. The term "population's optimal genes" refers to the reference genes that are the most suitable genotypes for ensuring survival within the population. These genes offer insights regarding the evolutionary trajectory of the population. It is necessary to delete the remaining individuals' information, including their ID, lifespan, and gene mutation rate, during the updating process.

**4.5. Experiment evaluation metrics**

In this study, we evaluated the performance of various LD measurements and haplotype block partitioning methods using a simulated dataset and experimentally validated HLA recombination hotspots. A total of nine joint methods were compared: CICC+CI, $r^2$+CI (PLINK), $D'$+CI (Haploview CI), CCC+CI, Haploview FGT, $D'$+SS (Haploview SS), CICC+SS, $r^2$+SS, and CCC+CI. These methods were selected due to their widespread use (e.g., PLINK and Haploview) and their ability to calculate haplotype block boundaries based on LD correlation results. While many alternative approaches exist for identifying block boundaries [21,22,73], most fail to directly determine boundaries informed by LD correlation outcomes, thereby lacking robust evidence for comparing LD measurement methods.

To better assess the differences between different methods in simulated data, we developed two different metrics: accuracy level and consistency level. Accuracy refers to the ratio of SNPs that are genuine components of a block. This metric is used to evaluate the precision of SNP allocation to blocks and determine whether SNPs are included in the same block.

Consistency level is another metric that needs to be considered, which quantifies the distribution of all identified SNPs in the synthetic block. The aim is to evaluate the alignment of blocks allocated differently on the boundary and impose a penalty on any deviation that occurs outside these limits. To facilitate a comprehensive assessment of the effectiveness of different methods, the F1 score was used.

$$F_1\ score = 2 * \frac{Accuracy * consistence\ level}{Accuracy + consistence\ level}$$

**4.6. Enrichment analysis**

To enable a more comprehensive comparison of CICC and $r^2$ across various disease datasets, we first identified candidate disease-related loci based on the core hypothesis that such loci show stronger LD in affected individuals, whereas LD is notably reduced in unaffected individuals. This reduction is thought to result from allelic variation in unaffected individuals, which may disrupt the co-adaptation or functional synergy among these loci. We then evaluated the biological relevance of the identified loci by examining what proportion had been independently validated in the GWAS Catalog (https://www.ebi.ac.uk/gwas/ ) as significantly associated with the corresponding diseases. This "proportion of functionally validated loci" served as our evaluation metric. By comparing these proportions between CICC and $r^2$, we assessed which LD measure offers greater biological interpretability in the context of real-world disease research.

For the enrichment analysis, we focused on complex diseases such as psoriasis, Alzheimer's disease (AD), and various mental disorders. Given the well-documented association between chromosome 6—particularly the HLA region—and both immune-related and psychiatric conditions, we prioritized our analysis on this chromosome.

In the analysis, for each dataset, we construct a SNP set $S = \{s_1, s_2, \ldots, s_n\}$. For any two distinct SNPs $s_i, s_j \in S\ (i \neq j)$, their pairwise correlation must satisfy the conditions $M_{case}(s_i, s_j) \geq l_{case}$ and $M_{control}(s_i, s_j) \leq l_{control}$. Here:

- $M$ denotes a measurement metric and can refer to either CICC or $r^2$.
- $M_{case}(s_i, s_j)$ denotes the correlation computed using metric M between SNPs $s_i$ and $s_j$ in the case dataset.
- $M_{case}(s_i, s_j)$ denotes the correlation computed using metric M between SNPs $s_i$ and $s_j$ in the control dataset.
- $l_{case}$ denotes the lower bound of correlation in the case dataset, and any value must exceed this threshold to meet the requirement.
- $l_{control}$ denotes the upper bound of correlation in the case dataset, and any value must be below this threshold to meet the requirement.

To identify SNP pairs that exhibit strong coordination in the case group but lack such coordinated structure in the control group, we defined distinct correlation thresholds for each dataset, denoted as $l_{case}$ and $l_{control}$. Given that $r^2$ values tend to be low overall, we adopted a relatively lenient threshold commonly used in GWAS studies and set $l_{case} = 0.6$. Unlike $r^2$, CICC differs systematically in both its measurement approach and distributional characteristics. Therefore, we adjusted the CICC threshold based on its empirical distribution and the network topology revealed through visualization, aiming to more accurately capture regions of high interdependence.

In the Psoriasis1 (ADO) dataset, the distribution of CICC values was slightly right-skewed (Fig. S12A). Based on this, we first applied standard filtering parameters and observed a marked shift in CICC values near 0.64 in the case group (Fig. S12B), suggesting a potential structural breakpoint. Consequently, we set the CICC threshold for the case group to $l_{case} = 0.64$. In contrast, no such inflection point was observed in the $r^2$ distribution, and the conventional threshold of $l_{case} = 0.6$ was retained. (Fig. S12C)

For the control group, where both CICC and $r^2$ exhibited similar distributions in the lower correlation range, we adopted a unified threshold of $l_{control}$ = 0.5, consistent with the standard demarcation for weak correlation. Overall, the correlation metrics in the Psoriasis1 (ADO) dataset exhibited mild skewness, with no notable outliers or structural anomalies, which supports the chosen thresholding strategy (Fig. S12A–C).

In the Psoriasis2 (GRU) dataset, the distribution of CICC values exhibited pronounced right skewness (Fig. S12D), indicating that high-correlation structures were confined to a limited subset of loci. To better capture these structures, we increased the threshold for the case group to $l_{case} = 0.7$ and lowered the threshold for the control group to $l_{control} = 0.2$. Although this parameter setting encompassed many SNPs, further analysis revealed that many loci exhibited low connectivity within the network, indicating that they did not form stable high-correlation structures with most other loci.

To address this, we examined the distribution of node co-occurrence frequencies across the network (Fig. S12E). A distinct inflection point was observed at a frequency of 7,300, suggesting a meaningful threshold for identifying structurally stable components. Accordingly, we implemented an additional filtering step, retaining only SNP pairs whose nodes co-occurred at least 7,300 times in the network. This refinement was intended to improve the robustness of the final selection.

The distribution of $r^2$ values remained consistently low, with no evident inflection point (Fig. S12F), justifying the continued use of the standard threshold of 0.6. In the control group, both CICC and $r^2$ exhibited markedly reduced correlation levels, with values concentrated in the very low range. Accordingly, a threshold of $l_{control} = 0.2$ was deemed appropriate and robust.

After preprocessing, the AD dataset contained over 920,000 SNP loci in both the case and control groups. To manage memory and I/O load, the data were partitioned into groups of 50,000 SNPs with 40,000 SNPs overlaps, resulting in 89 segment files for each group. To assess distributional consistency, we randomly selected 10% (9 files) and plotted their distributions (Fig. S12G and H), which showed overall consistency across segments.

Notably, CICC values showed clear changes near 0.2 and 0.8 (Fig. S12G), prompting us to set $l_{control}$= 0.2 and $l_{case}$= 0.8. We then merged the 89 filtered files after removing duplicates and plotted the frequency distribution of SNP occurrences (Fig. S12I). Since our goal was to identify loci that exhibit coordinated associations in the case group but disrupted coordination in controls, we aimed to retain SNPs with relatively uniform occurrence frequencies, avoiding overly skewed representation. A sharp change in the distribution was observed at a frequency of 5,000, and we therefore retained SNPs with co-occurrence frequencies greater than 5,000, yielding a final set of 1,572 loci.

The $r^2$ distribution in this dataset resembled that observed in Psoriasis2 (Fig. S12H), and we applied the same thresholds: $l_{control}$= 0.2 and $l_{case}$= 0.6. Based on the frequency distribution (Fig. S12J), a shift was observed at a frequency of 1,000, leading us to retain SNPs with occurrence counts above this threshold. This yielded a final set of 370 loci.

The Psychiatric Disorder (PD) dataset (PRJNA551447), which includes both case and control samples, was analyzed as PD. The distribution of CICC values in PD was left-skewed and displayed a bimodal pattern (Fig. S12K). Based on this, we set the control threshold to $l_{control}$= 0.3, while maintaining the case threshold at $l_{case} = 0.7$.

Compared to other datasets, $r^2$ values in PD were generally higher but showed no prominent fluctuations, supporting the continued use of the standard threshold (Fig. S12L). Initial filtering identified 945 loci. To further refine the selection, we examined the distribution of SNP co-occurrence frequencies (Fig. S12M) and observed a sharp drop at a frequency of 200. Applying this as a secondary threshold, we retained 13 high-frequency loci for downstream analysis.

Overall, the filtering strategy (Table 2B) in this dataset was guided by both the distributional properties of correlation metrics and the frequency of co-occurrence within the network, allowing for more accurate identification of stable, disease-associated structures.

*Competing interests*

The authors declare that they have no competing interests.

*Web resources*

1000G data is available from 1000 Genomes Project Consortium website (https://www.internationalgenome.org/).

HPRC data is available from the Human Pan-Genome Reference Consortium (https://humanpangenome.org/data/).

GAsP genomic data is from the GenomeAsia 100K project (https://browser.genomeasia100k.org/), which requires an application.

ADO and GRU are obtained from dbGaP (accession number: phs000019.v1.p1.c1)( https://www.ncbi.nlm.nih.gov/projects/gap/cgi-bin/study.cgi?study_id=phs000019.v1.p1 ).

Genotype data for Alzheimer's disease (AD) were acquired from the Alzheimer's Disease Neuroimaging Initiative (ADNI)( https://adni.loni.usc.edu/ ),

Genomic data pertaining to psychiatric diseases were obtained from the NCBI Sequence Read Archive, namely under accession numbers PRJNA551447(https://www.ncbi.nlm.nih.gov/bioproject/?term=PRJNA551447%20 ).

*Data availability*

Contact the authors for the simulated generation data used in this study.

*Acknowledgements*

The work was supported in part by funding from the Hong Kong RGC theme-based Strategic Target Grant Scheme (STG STG1/M-501/23-N), the Hong Kong Health and Medical Research Fund (HMRF grant 10211696), the Hong Kong Global STEM Professor Scheme, the Hong Kong Jockey Club Charity Trust, and PolyU Research Postgraduate Scholarship (PRPgS).

**Table 1. General information**

**(A) ALT allele frequency distribution in five populations**

| HLA region | Population | Mean | Variance | Max | Min |
|---|---|---|---|---|---|
| HLA Class I | AFR | 0.1627 | 0.059 | 0.9992 | 0.0008 |
|  | AMR | 0.1818 | 0.0593 | 0.9683 | 0.0014 |
|  | EAS | 0.1818 | 0.0625 | 0.999 | 0.001 |
|  | EUR | 0.1725 | 0.0543 | 0.9453 | 0.001 |
|  | SAS | 0.1916 | 0.0608 | 0.999 | 0.001 |
| HLA Class II | AFR | 0.2416 | 0.0675 | 0.9962 | 0.0008 |
|  | AMR | 0.2775 | 0.0721 | 0.9798 | 0.0014 |
|  | EAS | 0.2699 | 0.0727 | 0.995 | 0.001 |
|  | EUR | 0.2605 | 0.0621 | 0.9493 | 0.001 |
|  | SAS | 0.2515 | 0.0605 | 0.999 | 0.001 |

This table presents the frequency distribution of ALT across the HLA class I and HLA class II regions in five populations derived from the 1000 Genomes dataset, including metrics such as mean, variance, maximum value, and minimum value.

**(B) Gametic frequencies to allele frequencies at two loci**

|  |  | SNP2 | | Allele frequency |
|---|---|---|---|---|
|  |  | $B$ | $b$ |  |
| SNP1 | $A$ | $p_A p_B + D = p_{AB}$ | $p_A(1 - p_B) - D = p_{Ab}$ | $p_A$ |
|  | $a$ | $(1 - p_A)p_B - D = p_{aB}$ | $(1 - p_A)(1 - p_B) + D = p_{ab}$ | $p_a$ |
|  | Allele frequency | $p_B$ | $p_b$ | 1 |

**(C) LD measurements comparison in a toy case**

| SNP Info | | | Individuals | | | | | | | | | | | LD Measurement | | | | Marginal Probability | | | |
|---|---|---|---|---|---|---|---|---|---|---|---|---|---|---|---|---|---|---|---|---|---|
| SNP ID | REF | ALT | 1 | 2 | 3 | 4 | 5 | 6 | 7 | 8 | 9 | 10 | D' | CICC | $r^2$ | CCC | $p_A$ | $p_T$ | $p_C$ | $p_G$ |
| rs1 | A | T | A\|T | A\|A | A\|A | A\|A | A\|A | A\|A | A\|A | A\|A | A\|A | A\|A | 1 | 0.813 | 0.003 | 0.545 | 0.95 | 0.05 | 0 | 0 |
| rs2 | A | T | A\|A | A\|T | A\|A | A\|A | A\|A | A\|A | A\|A | A\|A | A\|A | A\|A | | | | | 0.95 | 0.05 | 0 | 0 |
| rs3 | C | G | C\|C | C\|C | C\|C | G\|G | G\|G | G\|G | G\|G | G\|G | G\|G | G\|G | 1 | 0 | 1 | 0.896 | 0 | 0 | 0.3 | 0.7 |
| rs4 | G | C | C\|C | C\|C | C\|C | G\|G | G\|G | G\|G | G\|G | G\|G | G\|G | G\|G | | | | | 0 | 0 | 0.3 | 0.7 |
| rs5 | C | G | C\|C | C\|C | C\|C | G\|G | G\|G | G\|G | G\|G | G\|G | G\|G | G\|G | 1 | 1 | 1 | 0.896 | 0 | 0 | 0.3 | 0.7 |
| rs6 | C | G | C\|C | C\|C | C\|C | G\|G | G\|G | G\|G | G\|G | G\|G | G\|G | G\|G | | | | | 0 | 0 | 0.3 | 0.7 |
| rs7 | C | G | C\|G | C\|G | C\|G | C\|G | C\|G | C\|G | G\|G | G\|G | G\|G | G\|G | 1 | 1 | 1 | 0.704 | 0 | 0 | 0.3 | 0.7 |
| rs8 | C | G | C\|G | C\|G | C\|G | C\|G | C\|G | C\|G | G\|G | G\|G | G\|G | G\|G | | | | | 0 | 0 | 0.3 | 0.7 |
| rs9 | C | G | C\|G | C\|G | C\|G | C\|G | C\|G | C\|G | G\|G | G\|G | G\|G | G\|G | 1 | 0.4 | 1 | 0.704 | 0 | 0 | 0.3 | 0.7 |
| rs10 | C | G | G\|C | G\|C | G\|C | G\|C | G\|C | G\|C | G\|G | G\|G | G\|G | G\|G | | | | | 0 | 0 | 0.3 | 0.7 |

This demo examines the marginal probability, gametes' mutation status and phasing information effects on linkage disequilibrium (LD) metrics—namely $D'$, CICC, $r^2$, and CCC—based on a dataset comprising 10 individuals and 10 SNPs. Each pair of SNPs is treated as a distinct group for analysis, resulting in a total of five comparative groups.

Comparing the first pair with the next four shows a strong dependence on current metrics, especially $D'$ and $r^2$, on marginal probabilities. Additionally, comparing the second and third groups reveals that existing metric formulas overlook base mutation scenarios. In diploid organisms like humans, every chromosome contains two copies with two alleles present. Upon comparing SNPs in the fourth pair and the fifth pair, it is evident that differences in haplotype composition do not affect $D'$, $r^2$, or CCC. Without haplotype genome information, both scenarios depicted in Fig 1A may be plausible and cannot be distinguished.

**(D) XOR example**

| SNP ID | REF | ALT | ind1 h1 | ind1 h2 | ind2 h1 | ind2 h2 | ind3 h1 | ind3 h2 | ind4 h1 | ind4 h2 | ind5 h1 | ind5 h2 | ind6 h1 | ind6 h2 | ind7 h1 | ind7 h2 | ind8 h1 | ind8 h2 | ind9 h1 | ind9 h2 | ind10 h1 | ind10 h2 | ALT.freq | $\Phi(X,Y)$ |
|---|---|---|---|---|---|---|---|---|---|---|---|---|---|---|---|---|---|---|---|---|---|---|---|---|
| rs1 | A | C | A | A | A | A | A | A | A | A | A | A | C | C | A | C | A | A | C | C | C | C | 0.35 | |
| rs2 | G | T | T | T | T | T | T | T | T | T | T | T | G | T | G | G | G | T | G | G | T | T | 0.7 | |
| | | | 0 | 0 | 0 | 0 | 0 | 0 | 0 | 0 | 0 | 0 | 0 | 1 | 1 | 0 | 1 | 0 | 0 | 0 | 1 | 1 | | 0.25 |
| rs3 | A | C | C | C | A | C | A | A | C | C | C | C | A | A | A | A | A | A | A | A | A | A | 0.35 | |
| rs4 | G | T | T | T | T | T | T | T | T | T | T | T | G | T | G | G | G | T | G | G | T | T | 0.7 | |
| | | | 1 | 1 | 0 | 1 | 0 | 0 | 1 | 1 | 1 | 1 | 1 | 0 | 1 | 1 | 1 | 0 | 1 | 1 | 0 | 0 | | 0.65 |

For a cohort of 10 individuals, each individual possesses two haplotype genomes. In this context, we utilize these 10 individuals as an example to illustrate the calculation of XOR and examine how the frequency of alternative alleles (ALTs) influences the XOR outcome. For rs1 in individual 1's h1, A is same as the REF, then $g(x_1) = 0$. For rs2 in individual 1's h1, T is same as the ALT, then $g(y_1) = 1$. Therefore, $g(x_1) \oplus g(y_1) = 0$.

**Table 2. Analysis result**

**(A) Summary of four measurements**

| Methods | Characters | Gene level | Haplotype level |
|---|---|---|---|
| $r^2$ | a) The overall value of $r^2$ is relatively small, and only the sites with a high degree of correlation will be distinctly presented.<br>b) It is not feasible to identify haplotypes or single genes of multiple genetic compositions. | It is feasible to identify sites with a high degree of correlation within a single gene that are appropriate for tagSNPs. | Hard to identify haplotypes. |
| $D'$ | a) $D'$ typically yields a higher correlation value throughout the entire range.<br>b) $D'$ is prone to identify extensive areas as having high LD, and in all four metrics, the probability of false positives is higher. | Highly correlated patterns within genes can be identified. | a) Suitable for identifying correlations among adjacent genes.<br>b) There is also a high correlation across gene regions. |
| CCC | CCC typically achieved correlations at the average correlation level. | Identify a limited number of genes. | Hard to identify haplotypes. |
| CICC | a) CICC is capable of recognizing both individual genes and haplotypes.<br>b) It fails to yield a globally high correlation value, thereby enabling more accurate identification of the sites with correlation.<br>c) Suitable for screening haplotypes, genes or variation sites of the entire genome. | Highly correlated patterns within genes can be identified. | Haplotypes can be identified. |

**(B) Comparison in enrichment analysis**

| Diseases | Measurements | $l_{case}$ | $l_{control}$ | Percentage | Found No. | Not Found No. |
|---|---|---|---|---|---|---|
| Psorisis1 (ADO) | CICC | 0.64 | 0.5 | 17.10% | 88 | 428 |
|  | $r^2$ | 0.6 | 0.5 | 16.90% | 34 | 167 |
| Psorisis2 (GRU) | CICC | 0.7 | 0.2 | 8.20% | 8 | 90 |
|  | $r^2$ | 0.6 | 0.2 | 0.00% | 0 | 4 |
| AD | CICC | 0.8 | 0.2 | 1.00% | 12 | 1144 |
|  | $r^2$ | 0.6 | 0.2 | 0.00% | 0 | 370 |
| PD | CICC | 0.7 | 0.3 | 0.00% | 0 | 39 |
|  | $r^2$ | 0.6 | 0.2 | 0.00% | 0 | 13 |

Each row shows the number of loci meeting the $l_{case}$ and $l_{control}$ conditions for each disease. "Found" indicates loci also reported in the GWAS Catalog (with percentages representing their proportions), whereas "Not Found" indicates loci not reported.

**(C) New regions with high LD identified by CICC in HLA Class I and HLA Class II**

|  | Region | Start ID | Start POS | End ID | End POS |
|---|---|---|---|---|---|
| HLA Class I | 1 | rs44012 | 31251365 | rs44348 | 31261638 |
|  | 2 | rs46604 | 31321100 | rs46748 | 31324805 |
|  | 3 | rs46748 | 31324805 | rs47036 | 31342147 |
| HLA Class II | 1 | rs57105 | 32375827 | rs57193 | 32383091 |
|  | 2 | rs57215 | 32384726 | rs57347 | 32390772 |
|  | 3 | rs57875 | 32421766 | rs58007 | 32427651 |
|  | 4 | rs58315 | 32448398 | rs58447 | 32459378 |
|  | 5 | rs59217 | 32488035 | rs61813 | 32723944 |
|  | 6 | rs63199 | 32779633 | rs63441 | 32788768 |
|  | 7 | rs64629 | 32883221 | rs65113 | 32932967 |

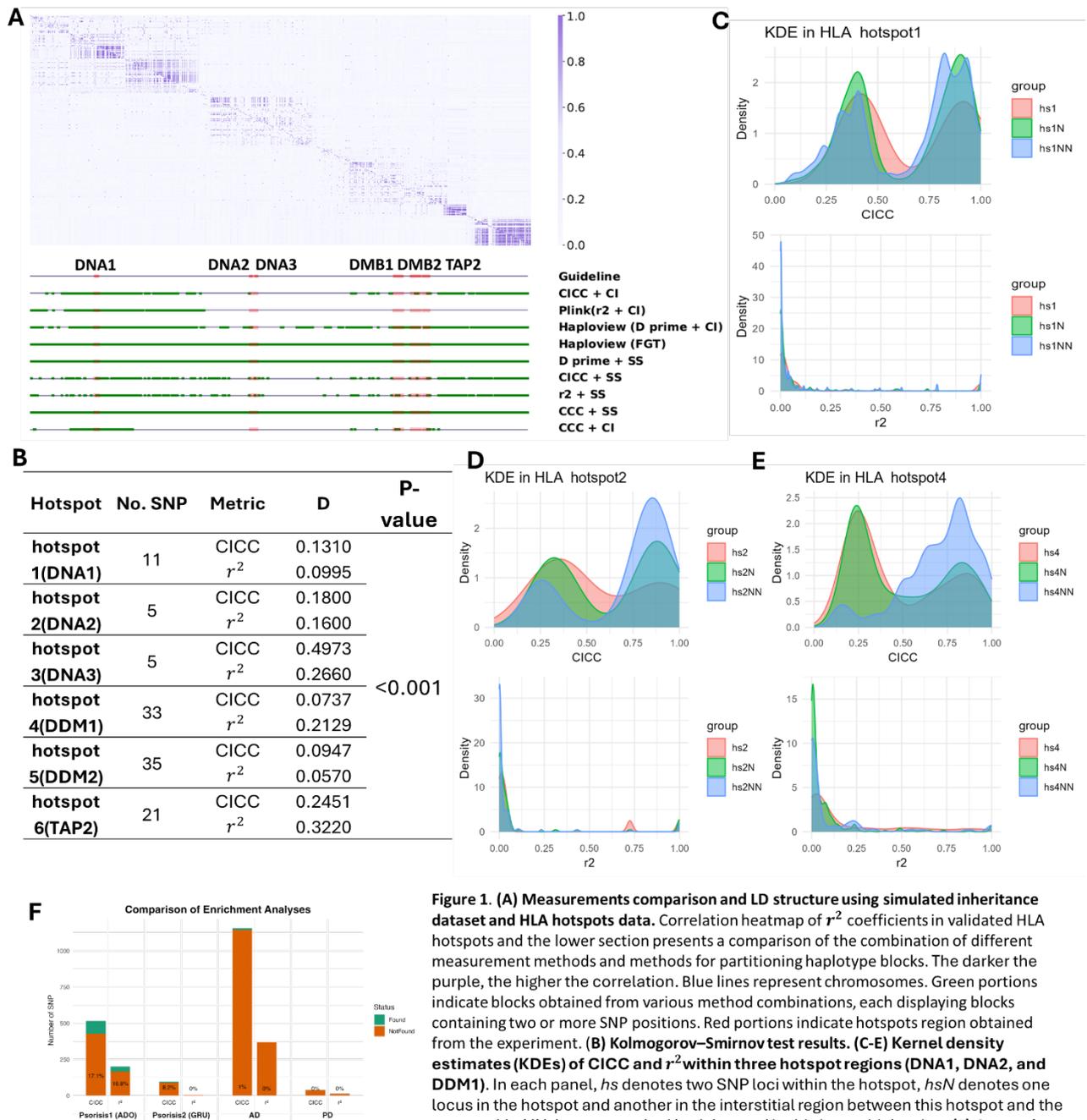

Figure 1. (A) Measurements comparison and LD structure using simulated inheritance dataset and HLA hotspots data. Correlation heatmap of $r^2$ coefficients in validated HLA hotspots and the lower section presents a comparison of the combination of different measurement methods and methods for partitioning haplotype blocks. The darker the purple, the higher the correlation. Blue lines represent chromosomes. Green portions indicate blocks obtained from various method combinations, each displaying blocks containing two or more SNP positions. Red portions indicate hotspots region obtained from the experiment. (B) Kolmogorov–Smirnov test results. (C-E) Kernel density estimates (KDEs) of CICC and $r^2$ within three hotspot regions (DNA1, DNA2, and DDM1). In each panel, hs denotes two SNP loci within the hotspot, hsN denotes one locus in the hotspot and another in the interstitial region between this hotspot and the next, and hsNN denotes two loci both located in this interstitial region. (F) Comparison in enrichment analysis. Each bar shows the number of loci meeting the $l_{case}$ and $l_{control}$ conditions for each disease (x-axis); green indicates those also reported in the GWAS Catalog (numbers show proportions), and orange indicates those not reported.

**Figure 1.** (A) Measurements comparison and LD structure using simulated inheritance dataset and HLA hotspots data. Correlation heatmap of $r^2$ coefficients in validated HLA hotspots and the lower section presents a comparison of the combination of different measurement methods and methods for partitioning haplotype blocks. The darker the purple, the higher the correlation. Blue lines represent chromosomes. Green portions indicate blocks obtained from various method combinations, each displaying blocks containing two or more SNP positions. Red portions indicate hotspots region obtained from the experiment. (B) Kolmogorov–Smirnov test results. (C-E) Kernel density estimates (KDEs) of CICC and $r^2$ within three hotspot regions (DNA1, DNA2, and DDM1). In each panel, hs denotes two SNP loci within the hotspot, hsN denotes one locus in the hotspot and another in the interstitial region between this hotspot and the next, and hsNN denotes two loci

both located in this interstitial region. (F) Comparison in enrichment analysis. Each bar shows the number of loci meeting the $l_{case}$ and $l_{control}$ conditions for each disease (x-axis); green indicates those also reported in the GWAS Catalog (numbers show proportions), and orange indicates those not reported.

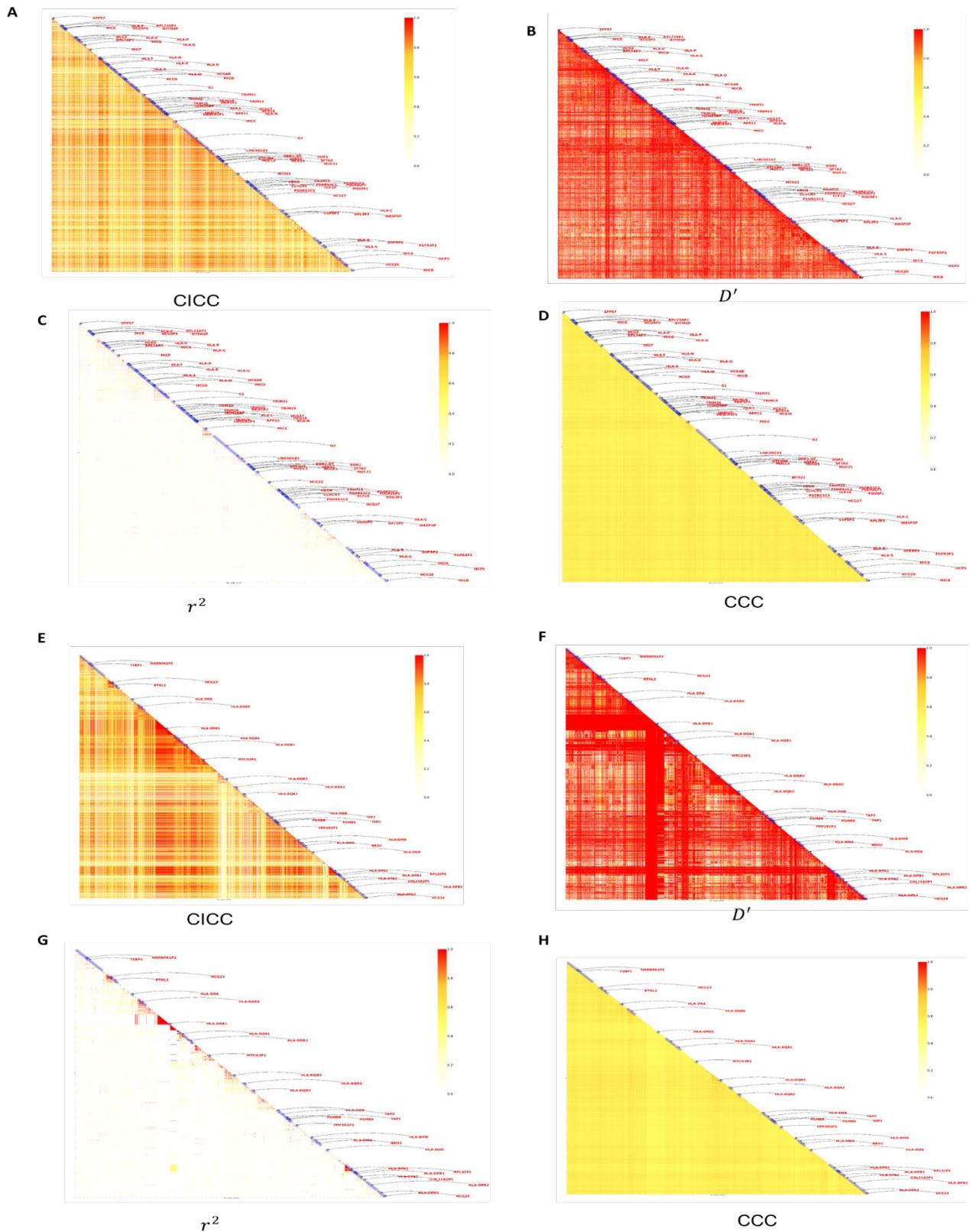

**Figure 2. LD structure of HLA class I and class II based on haplotype genomes.** (A)-(D) is the HLA class I LD structure obtained from CICC, $D'$, $r^2$ and CCC, respectively. (E)-(H) is the HLA class II LD structure obtained

from CICC, $D'$, $r^2$ and CCC, respectively. The redder the color, the higher the correlation. G1 contains HCG8, C6orf12, HLA-J, ETF1P1, ZNRD1, PPP1R11, and RNF39. G2 contains TMPOP1, SUCLA2P, RANP1, HLA-E, GNL1, PRR3, ABCF1, PPP1R10, MRPS18B, ATAT1, PTMAP1, PTMAP1, C6orf136, DHX16, PPP1R18, NRM, RPL7P4, MDC1, TUBB, FLOT1, IER3, and HCG20.

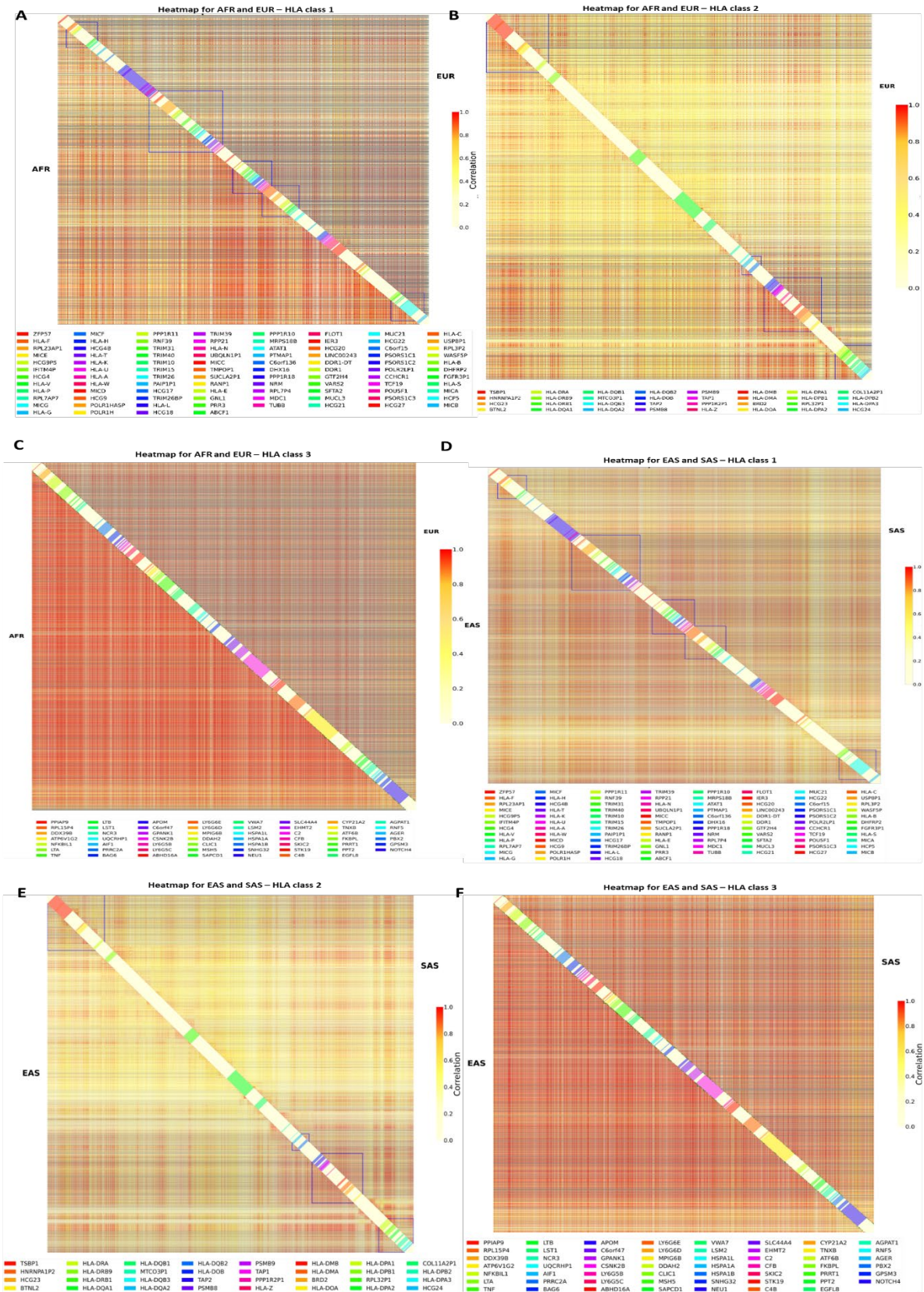

**Figure 3. Comparison of HLA LD structure.** (A-C) HLA structure of AFR and EUR in HLA class I, II and III. (D-F)

HLA structure of EAS and SAS in HLA class I, II and III. The upper and lower triangles of each square represent SNPs from the two respective regions. Gray bars indicate loci that are absent in one population but present in the other. Blue outlined boxes denote boundaries of strongly correlated regions consistently identified across different populations. The redder the color, the higher the correlation.